\documentclass{osa-article}

\journal{oe}

\newcommand{\sket}[1]{{\ensuremath{\lvert#1\rangle}}}
\newcommand{\lket}[1]{{\ensuremath{\left\lvert#1\right\rangle}}}
\newcommand{\ket}[1]{\if@display\lket{#1}\else\sket{#1}\fi}

\newcommand{\sbra}[1]{{\ensuremath{\langle#1\rvert}}}
\newcommand{\lbra}[1]{{\ensuremath{\left\langle#1\right\rvert}}}
\newcommand{\bra}[1]{\if@display\lbra{#1}\else\sbra{#1}\fi}

\newcommand{\sbraket}[2]{{\ensuremath{\langle#1\rvert#2\rangle}}}
\newcommand{\lbraket}[2]{{\ensuremath{\left\langle#1\!\left\rvert\vphantom{#1}#2\right.\!\right\rangle}}}
\newcommand{\braket}[2]{\if@display\lbraket{#1}{#2}\else\sbraket{#1}{#2}\fi}

\newcommand{\sketbra}[2]{{\ensuremath{\lvert #1\rangle\!\langle #2\rvert}}}
\newcommand{\lketbra}[2]{{\ensuremath{\left\lvert #1\right\rangle\!\!\left\langle #2\right\rvert}}}
\newcommand{\ketbra}[2]{\if@display\lketbra{#1}{#2}\else\sketbra{#1}{#2}\fi}

\usepackage{soul}
\usepackage{changes}

\begin{document}

\title{High-Rate Photon Pairs and Sequential Time-Bin Entanglement with $\mathbf {Si_3N _4}$ Ring Microresonators}

\author{Farid Samara,\authormark{1} Anthony Martin,\authormark{1} Claire Autebert,\authormark{1}Maxim Karpov,\authormark{2} Tobias J. Kippenberg,\authormark{2}  Hugo Zbinden,\authormark{1} and Rob Thew\authormark{1*}}

\address{\authormark{1}Group of Applied Physics, University of Geneva, Geneva, Switzerland\\
\authormark{2}Swiss Federal Institute of Technology Lausanne, CH-1015, Switzerland\\
}

\email{\authormark{*}robert.thew@unige.ch} 



\begin{abstract}
Integrated photonics is increasing in importance for compact, robust, and scalable enabling quantum technologies. This is particularly interesting for developing quantum communication networks, where resources need to be deployed in the field. We exploit photonic chip-based $ \rm Si_3 N_4$ ring microresonators to realise a photon pair source with low-loss, high-noise suppression and coincidence rates of 80$\times 10^3$\,s$^{-1}$. A simple photonic noise characterisation technique is presented that distinguishes linear and nonlinear contributions useful for system design and optimisation. We then demonstrate an all-fibre 750\,MHz clock-rate sequential Time-Bin entanglement scheme with raw interference visibilities $>$\,98\,\%. 
\end{abstract}

\section{Introduction}
Integrated photonics allows one to realise optical devices that are robust, compact and scalable, thus motivating the increasing interest in research and development of integrated optical components for both the classical and quantum communication domains~\cite{politi2009integrated,Tanzilli2011}. 
In particular, integrated sources of single and entangled photons at telecommunication wavelengths~\cite{o2009photonic,eisaman2011invited} are an essential resource for the field of quantum communication~\cite{Gisin2007}.

By exploiting the nonlinear processes of spontaneous parametric down-conversion (SPDC) or spontaneous four-wave mixing (SFWM), such photon sources have been demonstrated in a variety of integrated photonic structures, including periodically poled nonlinear crystals \cite{Tanzilli2011}, strip waveguides~\cite{takesue2007entanglement,zhang2016correlated}, disks~\cite{fortsch2013versatile,lu2016heralding} and ring resonators~\cite{mazeas2016high,pasquazi2017micro}. 
Among the different approaches, a particularly promising one is the use of SFWM in microring resonator (MRR) cavities. Indeed, a significant practical advantage is provided due to the field enhancement offered by the MRR cavity, which reduces the required pump power to the milliwatt or even the microwatt regime~\cite{Savanier2016a}. 

MRR devices generate correlated photon pairs distributed over a frequency-comb, corresponding to the cavity resonances around the pump wavelength. This frequency comb structure can be matched to different DWDM channels to multiplex the entanglement~\cite{mazeas2016high,fujiwara2017wavelength} or to generate high-dimensional frequency entangled quantum states~\cite{imany201850,kues2017chip,jaramillo2017persistent} which can be used for quantum computation~\cite{lukens2017frequency,babazadeh2017high}. Additionally, optical cavities, such as MRR, can produce inherently narrow bandwidth photons~\cite{ou1999cavity,akbari2016third} without the need for narrowband filters, which can introduce unwanted loss. SFWM in MRR structures have demonstrated photon bandwidths down to 30\,MHz~\cite{ramelow2015silicon}. Such narrow photon bandwidths are well adapted to the acceptance bandwidth of quantum memories~\cite{Simon2010}, thus they are important in the application of quantum networks.

Silicon photonics is a commonly employed platform for integrated photon pair sources due to its high third-order non-linearity and CMOS-compatibility, with numerous silicon-MRR photon pair sources demonstrated so far \cite{faruque2018chip,Savanier2016a,Hemsley2016,Leuthold2010}. Nevertheless, silicon suffers from significant two-photon absorption at telecom wavelength due to a relatively low bandgap \cite{pasquazi2017micro}. This effect is even more pronounced in MRR due to the field enhancement associated with the cavity, resulting in $Q$-factor and photon pair generation limits. In contrast, silicon nitride ($ \rm Si_3 N_4$) and high-index doped silica (Hydex) \cite{levy2010cmos} provide promising alternative platforms due to their CMOS-compatibility and absence of two-photon absorption in the telecommunication band, with several recent demonstrations of MRR based photon pair sources for both platforms~\cite{Reimer:14,Moss2013,reimer2016generation,ramelow2015silicon,imany201850}. In particular, we focus on the $ \rm Si_3 N_4$  platform, which due to recent advances in the fabrication of ultra-low-loss waveguides~\cite{pfeiffer2016photonic,Pfeiffer18}, flexible dispersion engineering~\cite{pfeiffer2017octave} and fiber-to-chip coupling techniques~\cite{liu2018double}, has become one of the key platforms for nonlinear integrated photonics. A prominent example of its application are photonic-chip-based soliton microcombs~\cite{Kippenberg18,brasch2016photonic}, which have enabled the generation of broadband coherent optical combs with high repetition rates in fully integrated devices amenable for wafer-scale manufacturing. 

In this paper, we study an integrated photon-pair source based on SFWM in $ \rm Si_3 N_4$ MRR. This highlights the advantages of using $ \rm Si_3 N_4$ as well as introducing a novel approach for studying unwanted photonic noise in SFWM sources in general. High photon pair detection rates were achieved thanks to efficient noise filtering. Additionally, we exploit a $ \rm Si_3 N_4$ MRR to demonstrate a high repetition rate sequential Time-Bin scheme \cite{Zhang:08} well suited for quantum communication applications.

\section{Experimental Set-up}
\label{setupS}
\begin{figure}[htbp]
	\center\includegraphics[width=\textwidth]{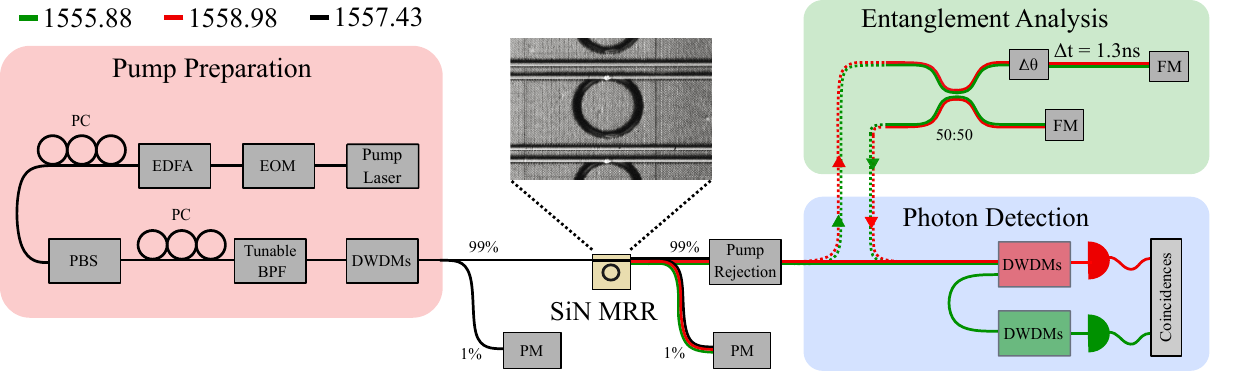}
	\caption{Experimental schematic. On the left is the pump laser preparation which is then injected into the $ \rm Si_3 N_4$ MRR. On the right, we have the set-up for characterising the source and a folded Michelson interferometer for analysing the sequential Time-Bin entanglement. EOM: electro-optical modulator; EDFA: erbium doped fibre amplifier. PC: polarisation controller. PBS: fibre polarising beam-splitter. BPF: Band-Pass Filter. PM: power meter. FM: Faraday mirror.}
	\label{setup}
\end{figure}

Fig.~\ref{setup} illustrates the set-up used for the photon pair generation and detection, in addition to the set-up for sequential Time-Bin entanglement analysis. The MRR under test has a $Q$-factor of $4.6\,\times\,10^5$, which gives a good trade-off between brightness and stability: higher $Q$-factors lead to brighter sources but narrower resonance linewidths, and hence, are more challenging to maintain on-resonance without sophisticated stabilisation schemes \cite{Reimer:14,Savanier2015}. This also results in relatively short coherence times allowing for high repetition rates, e.g. for the sequential Time-Bin entanglement scheme. For the moderate $Q$-factor devices used here, thermal stabilisation of the chip was enough to ensure on-resonance operation. The free spectral range (FSR) of the device is 192.37\,GHz, which is sufficiently close to the standard 200\,GHz DWDM channel spacing for low-loss filtering. The resonances structure can be thermally tuned (-2.75\,GHz/K) to achieve the correct matching with the DWDM channels.

In the first instance, we use a continuous-wave (CW) pump laser (DL100 Toptica) at a wavelength of $\lambda\,=\,1557.43\,{\rm nm}$ corresponding to the middle of the ITU channel 25. The pump can be modulated by an electro-optical modulator (EOM), thus giving rise to pulses which are used to generate sequential Time-Bin entanglement~\cite{brendel1999pulsed}. The pulse width and the repetition rate can be fixed by choosing the corresponding radio frequency (RF) signal that drives the EOM. Subsequently, we use an erbium-doped fibre amplifier (EDFA) to achieve the desired amount of power. A set of polarisation controllers and a polarising beam splitter (PBS) are used to control the polarisation coupled into the waveguide. To remove the amplified spontaneous emission of the pump (laser + EDFA), a pump filtering stage is employed, consisting of a tunable bandpass filter (25~GHz at FWHM) and two 100~GHz DWDM filters which give a pump isolation of 135~dB.

Light is coupled in and out of the chip using an anti-reflection (AR) coated lensed fibre with a spot-size diameter of 5\,$\mu$m. The chip to lensed-fibre coupling losses at both facets are continuously monitored by collecting 1\,\% of the light at the input and the output of the chip; on average, the input to output losses are -6~dB. If we neglect the propagation loss inside the bus waveguide, and assume equal coupling losses for the input and output facets, then we get -3~dB of coupling loss per facet. At the output of the chip, a set of DWDMs is employed to reject the pump and select the photons of interest.  

The detection scheme is based on superconducting nanowire single-photon detectors (SNSPDs)~\cite{Misael2018} with a detection efficiency of 80\,\% and dark count rate of 40\,s$^{-1}$ whose signals are sent to a time-to-digital-converter (IDQ-ID800) to record the coincidences histogram.

\section{Photon Pair Source Characterisation}
\label{Characterisation}
In the following we characterise the source, in particular, we present a novel approach to determine various photonic sources of noise in the system, as well as pair generation rates and the purity of the emitted photons.

\subsection{Photonic Noise On and Off the Chip}
\label{noise}

Ideally, the source should generate only correlated photon pairs, however, in such an experiment there are several sources of unwanted photonic noise that can degrade their quality or the subsequent entanglement measurement. In our experiment, we can identify two main categories of photonic noise: the first is related to the set-up without the $ \rm Si_3 N_4$ MRR chip, while the second is due to the MRR chip itself, and in particular, only when the pump wavelength is tuned on-resonance. In the following, we refer to all types of uncorrelated photons as "photonic noise" and we first try to understand their origins, and subsequently, to minimise their contribution.

\begin{figure}[htbp]
	\includegraphics[width=\linewidth]{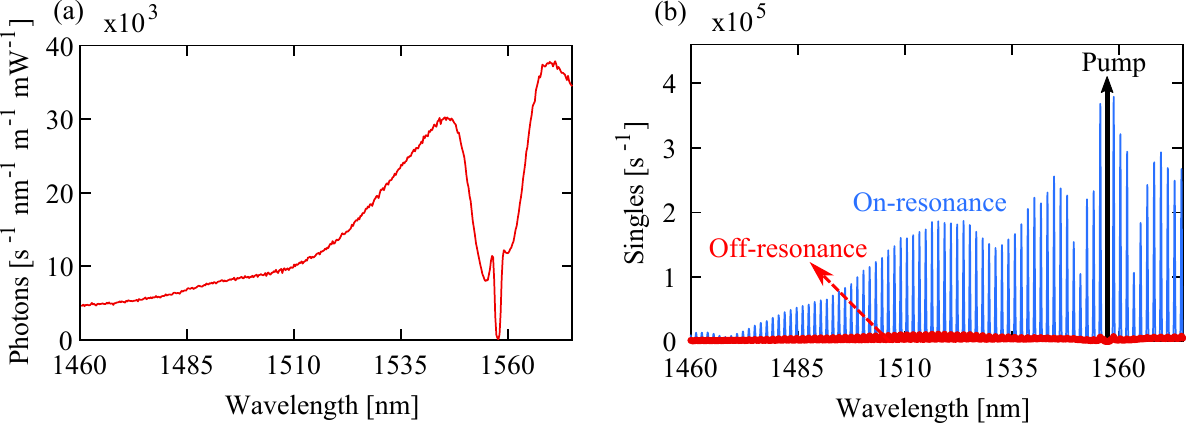}
	
	\caption{(a) Spontaneous Raman scattering generated by the set-up without the MRR. It is possible to observe two dips around the pump wavelength of 1557.43~nm; A narrow dip due to the pump rejection filters, and a wide dip due to spontaneous Raman scattering in the fibres. (b) The spectral response of the MRR for both on- and off-resonance cases.}
	\label{Group1}
\end{figure}
In the first case, we investigate the noise introduced by the experimental set-up by bypassing the MRR device. To characterise the spectrum of the photonic noise we consider either the signal or idler path, and replace the DWDM filters of the photon detection stage (see Fig.~\ref{setup}) with a tunable bandpass filter with a bandwidth of 25\,GHz combined with a set of 200\,GHz notch filters (not shown in  Fig.~\ref{setup}). The combination of these filters provides a total pump rejection of 110\,dB. The noise spectrum is presented in Fig.~\ref{Group1}.a. We observe that the pump laser (around 1557\,nm) residue is negligible, however, elsewhere, there is a broad spectral envelope resembling that of typical Raman scattering in fibre~\cite{Eraerds2010}. By varying the pump power and fibre length we determine a linear scaling, in keeping with the hypothesis of Raman generated noise. To mitigate this noise, we shortened the fibres between the pump cleaning and pump rejection filters down to a minimum practical length of around 50\,cm in total. 

Having studied the photonic noise related to the fibre elements of our set-up, we now turn our attention to studying the extra contribution that is introduced by the MRR $ \rm Si_3 N_4$ chip. For this, we insert the MRR chip back in the set-up and observe the single photon counts when the pump wavelength is tuned both \textit{off} and \textit{on} the cavity resonance wavelength, as shown in Fig.~\ref{Group1}.b.  We can clearly see the comb structure associated with the MRR emission, which spans around 60 frequency comb lines on the blue-shifted side of the pump. A similar spectrum exists on the red-shifted side, however, this is outside the range of our tunable filter. 

\begin{figure}[!htbp]
	\includegraphics[width=\linewidth]{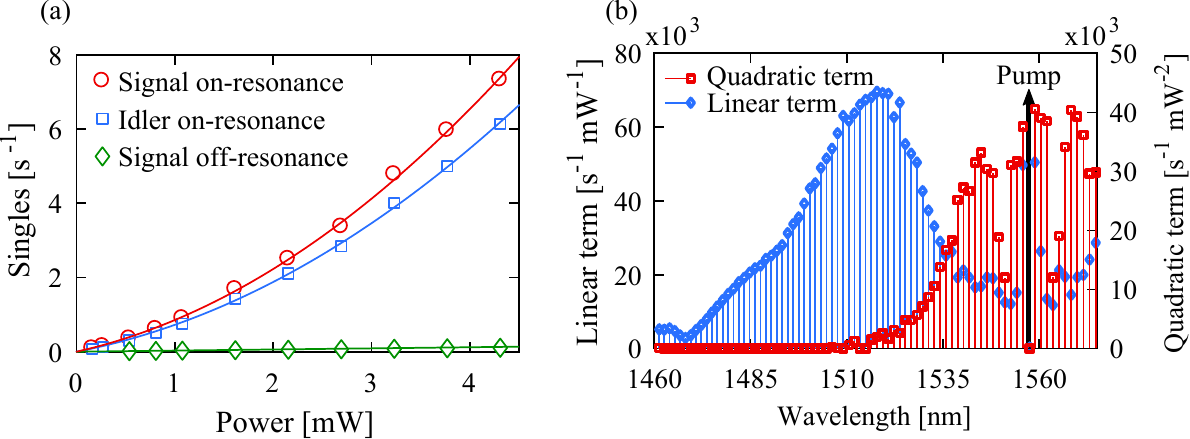}
\caption{(a) Detected photons (Singles) as a function of injected laser pump power for both the \textit{off} and \textit{on} resonance cases. For the off-resonance case, only the singles from the signal wavelength are shown. (b) The singles as a function of power for each resonant line is separated into its linear and a quadratic terms.} \label{Group2}
\end{figure}

To distinguish the contributions of photonic noise and the correlated photon pairs of interest, i.e. SFWM, we measure the spectral response for different pump powers at the input of the MRR. Fig.~\ref{Group2}.a shows the number of detections as a function of pump laser power for the two frequency channels neighbouring the pump, i.e. the target signal and idler wavelengths. We can then fit this data with a function of the form: $a\,P + b\,P^2$, where $P$ correspond to the injected pump power. The photonic noise due to the linear process can be associated with the coefficient $a$, and the quadratic contribution of photons generated by SFWM with $b$. Overall, the off-resonance linear noise contribution is comparable to the case with no chip, around \,$a=2.5\times\,10^3$\,s$^{-1}$\,mW$^{-1}$, however, on-resonance the noise significantly increases to \,$a=26\times\,10^3$\,s$^{-1}$\,mW$^{-1}$. In comparison, we find \,$b=59\times\,10^3$\,s$^{-2}$\,mW$^{-2}$ for the quadratic contribution in the signal channel. The on-resonance linear contribution indicates that there is extra photonic noise generated by the MRR chip. 

In Fig.~\ref{Group2}.b. we extend this analysis to the individual frequency channels, repeatedly measuring each channel for varying pump powers and extracting the linear and quadratic contributions. We see that while the SFWM term is high for the resonances comb adjacent to the pump wavelength, there is also the presence of significant linear noise. This linear noise appears only for the on-resonance case, thus indicating that it is being generated by the MRR cavity, possibly due to self-phase modulation of the pump wavelength.

The spectral shape of the linear noise contribution resembles that reported for fibre~\cite{Eraerds2010}, however, it is not obvious that this can be completely attributed to a Raman process. One also needs to consider the role of the SiO$_2$ substrate and tighter confinement of the MRR. Recently, Raman-like spectra have been, at least partially, attributed to thermo-refractive noise in amorphous materials such as $ \rm Si_3 N_4$ waveguides~\cite{LeThomas18}. A better understanding of the origins could further assist in optimising the fabrication of these devices and their exploitation as low-noise photon pair sources. Nonetheless, this way to analyse the photonic noise provides a practical approach to identify the best DWDM channels to maximise the signal to noise ratio and CAR.

\subsection{Photon Pair Generation}
\label{Pair}
To characterise the photon pair generation we use the analysis setup illustrated in the lower right corner of Fig.~\ref{setup}. Now that there is no need to tune filters as for the noise analysis, the signal, and idler photons are collected with two band-pass (200\,GHz) DWDMs in the  frequency channels immediately around the pump at 1558.98\,nm (ITU23) and 1555.88\,nm (ITU27), respectively, with an additional notch DWDM filter and a fibre Bragg grating, which achieves a pump rejection of 145\,dB with a signal and idler isolation of 100\,dB.

\begin{figure}[!htbp]
	\includegraphics[width=\linewidth]{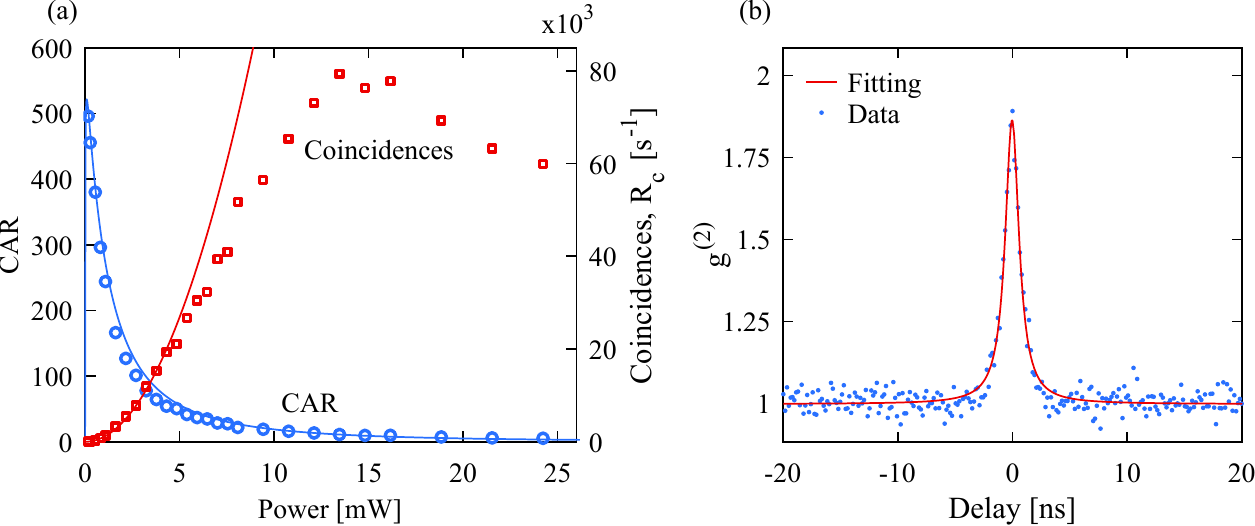}
	\caption{(a) Coincidence count rate $R_c$ and coincidence-to-accidental ratio CAR as a function of the pump laser power. The predicted CAR and the quadratic fitting of the coincidences at low power (before detector saturation) are given by solid lines. (b) The auto-correlation coincidence histogram, from which we can extract the purity of the generated photons.}
	\label{Group3}
\end{figure}

The presence of correlated photon pairs was demonstrated by acquiring the coincidence histogram between the signal and idler detections and observing the coincidence peak at zero delay. The coincidence count rates and the coincidences-to-accidental ratio (CAR) as a function of the injected pump power are shown in Fig.~\ref{Group3}.a.
We see that for pump powers below 5\,mW, the coincidence count rate scales with the square of the pump power, thus confirming that the generated pairs are coming from the SFWM process. However, as the pump power is further increased, we enter the count rate saturation region of the SNSPDs where the effective detection efficiency begins to drop as the detectors do not have time to recover their full efficiency. This explains the deviation of the coincidence count rate from the quadratic scaling for pump powers above 5\,mW. This is exacerbated due to the relatively high linear photonic noise that produces singles but no coincidence counts. Nonetheless, with a pump power of 13.5\,mW, we obtain a maximum photon pair detection rate of 80\,$\times\,10^3$\,s$^{-1}$ with a corresponding CAR of 12.3.

To understand the limiting factors on the CAR, also shown in Fig.~\ref{Group3}.a., we calculated the expected behaviour of the CAR as a function of the laser pump power (see \cite{clausen2014source} for example) based on the photonic noise found in Sec.~\ref{noise}. The CAR prediction agrees with the experimental data. It shows a reduction of CAR at high pump power due to double pair generation. At low pump power, the CAR is mainly limited by the detector's dark counts. The highest CAR value of 495 was obtained by injecting 160\,$\mu W$ of pump power, however, this corresponds to a relatively low coincidence detection rate of 29.5\,s$^{-1}$. It is worth mentioning that these numbers indicate a range of operation, which in any case would need to be optimised for a given experiment.

We can relate the pair generation rate (PGR) to the singles for the signal $S_s$, idler $S_i$ and coincidences $R_c$ by $PGR = S_iS_s/R_c$~\cite{tanzilli2001highly}. By considering only the singles due to SFWM, i.e. the quadratic terms from the fittings in Fig.~\ref{Group2}.a. and the $R_c$ from the fitting of Fig.~\ref{Group3}.a., we estimate a PGR of $5.2\,\times\,10^5$\,s$^{-1}$\,mW$^{-2}$. The photons bandwidth and coherence time can be derived from the coincidence histogram, or the resonance line-width, which gives 210\,MHz and 760\,ps respectively, giving a source brightness of $2.5\,\times\,10^3$\,s$^{-1}$\,MHz$^{-1}$\,mW$^{-2}$. One should note that the brightness should only be compared to $\chi^{(3)}$ sources and for MRR this is highly dependent on the $Q$-factor of the resonator, for which we chose here a relatively low value. For example, Ramelow \textit{et. al.}~\cite{ramelow2015silicon} obtained a PGR of $3.9\,\times\,10^6$\,s$^{-1}$\,mW$^{-2}$ for 90\,MHz photons from a MRR device with a $Q=2\times 10^6$.
Finally, the total loss in each of the signal and idler paths can be estimated from the ratio between the measured singles and coincidence counts~\cite{tanzilli2001highly}, which give -13.05~dB and -13.84~dB for signal and idler respectively. This total loss includes the propagation losses inside the MRR, the coupling loss, filter losses, and the finite efficiency of the detectors.


\subsection{Photon Purity}
\label{purity}

Pure, or spectrally uncorrelated, photon-pairs are essential for the realisation of efficient multi-photon experiments such as teleportation or entanglement swapping. By generating pure photons, there is no need to use extra, and typically lossy, spectral filtering to ensure that photons from different sources are indistinguishable - having a high HOM dip visibility~\cite{Bruno14}. n.b. the filter bandwidths used here are much larger than that of the photons such that the photons are distributed into their respective wavelength (DWDM) channels, but not spectrally filtered.
The state purity can be determined by measuring the second-order auto-correlation function - $g^{(2)}(0)$ - in a Hanbury Brown and Twiss like experiment~\cite{Tapster1998} by sending the signal (or idler) photons onto a 50:50 beam-splitter and measuring the coincidence detections between the outputs~\cite{Christ2011}. For a pure state, we expect a $g^{(2)}(0)$ close to 2, which corresponds to thermal statistics typical of a single mode. When the number of modes increases $g^{(2)}(0)$ tends to 1, which is the signature of a Poissonian distribution. The results of the $g^{(2)}(0)$ measurement is reported in Fig.~\ref{Group3}.b. The data was fitted with a Lorentzian function, from which we calculated $g^{(2)}(0)$ of $1.86\pm0.07$ (taken as the ratio between the coincidences at zero-delay and the background). The photon purity can also be quantified by the Schmidt number $n$~\cite{Christ2011,Clausen_2014}, which is related to the auto-correlation function by $g^{(2)}(0) = 1 + 1/n$. From a $g^{(2)}(0)$ of $1.86\pm0.07$ we find $n = 1.16\pm0.11$.

\section{Sequential Time-Bin Entanglement}
\label{STBE}

The advantage of Time-Bin entanglement, compared to Energy-Time, is the ability to synchronise independent sources for more complex communication tasks while keeping the advantage of phase encoding. We start by using an electro-optical modulator (EOM), as shown in Fig.\ref{setup}, to pulse the pump laser such that photon pairs are created in superpositions of different times.  The CW laser ensures that coherence is maintained over these time-bins such that we generate states of the form, 
\begin{equation}
\sket{\psi}  \propto \sket{t_i,t_{i}}_{\omega_1 \omega_2} + e^{i\phi} \sket{t_{i+1},t_{i+1}}_{\omega_1 \omega_2}.\label{entTB}
\end{equation}
Rather than sending each laser pulse through a ``Time-Bin interferometer'', we use each sequential pair of laser pulses to define the time bins, such that the laser pulse rate can be fully exploited. To characterise the entanglement we need to pass the pairs through imbalanced interferometers with a path-length difference $\Delta t$ corresponding to the laser's periodicity. In this experiment we use a device with a slightly lower $Q$-factor of $1.1 \times 10^5$ to ensure that the coherence time of the generated photons $\tau_{c} \approx 180\,$ps $\ll  \Delta t$. This then allows us to operate with a clock rate for the pulsed laser of 750\,MHz, i.e. with $\Delta t= 1.33~ns$.

Typically, in such an experiment, each photon of a pair is sent to an individual interferometer, however, for simply certifying the entanglement it is sufficient to send both photons of the pair into an imbalanced Michelson fiber interferometer, as shown in the top right of Fig.\ref{setup}. The interferometer is temperature stabilised and a piezo is used to vary the relative phase between the two arms ($\Delta\phi$). The Time-Bin interferometer introduces -6\,dB of extra loss, with -3\,dB due to the beam splitter. 



\begin{figure}[!htbp]
	\centering\includegraphics{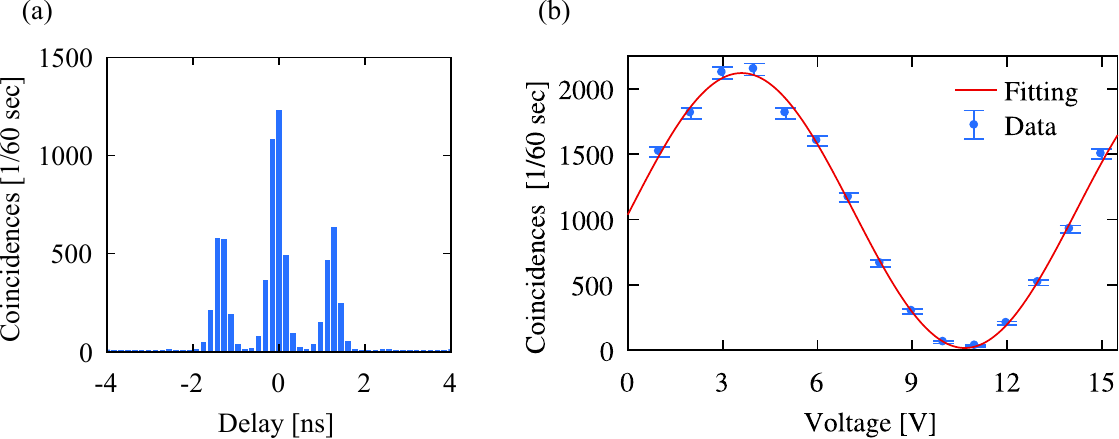}
	\caption{(a) The average coincidence histogram. (b)  Coincidence counts in the central and side histogram peaks as a function of the phase - voltage on the piezo.}
	\label{Group4}
\end{figure}
In Fig.~\ref{Group4}.a. we can see the characteristic three temporal peaks clearly defined in the coincidence histogram, where we post-select only the central one which corresponds to a measurement of the state of Eq.\ref{entTB}. 
The coincidence counts in the central peak are integrated for 60\,s for each phase setting giving rise to the interference curve in Fig.~\ref{Group4}.b. From the fitting, we extracted a raw visibility of $98.29\pm 0.20$\,\% and a net visibility of $99.96\pm 0.03$\,\%. In contrast, no interference is observed in the coincidence counts of the side peaks. These results clearly demonstrate a high degree of entanglement. We see that the system achieves a coincidence rate of around $R_c \sim16\,$s$^{-1}$ for 10~$mW$ of injected pump power.

\section{Discussion}

We have demonstrated a $ \rm Si_3 N_4$ MRR scheme for generating photon pairs and sequential Time-Bin entanglement that has relatively low losses, with excellent pump extinction and pair isolation. The noise analysis developed here provides a useful tool to identify frequency channel pairs that have the best CAR. While spurious photonic noise from the associated set-up, principally in the fibres, was largely suppressed, photonic noise from the $ \rm Si_3 N_4$ MRR remains and warrants further investigation to better understand its origins and potential mitigation solutions. Nonetheless, the resulting sequential Time-Bin entanglement (raw) visibility $>$98\,\% and pair coincidence rates of $80\times 10^3$\,s$^{-1}$ demonstrate this is not a fundamentally limiting factor. The performance of these devices is increasingly comparable with more mature technologies, such as PPLN \cite{Tanzilli2011}, with the benefit of using a telecom wavelength pump that is advantageous for distributed quantum communication scenarios.

The recent progress on high-$Q$ MRR in $ \rm Si_3 N_4$~\cite{Pfeiffer18} could also provide an avenue for more complex quantum communication tasks such as multiphoton entanglement swapping experiments, or coupling to quantum memories.  An area where significant impact could be made is in better mode matching between the fibre and bus waveguide for fibre pigtailing, which would be important to increase both the coupling and heralding efficiencies. The device and system performance could also be further improved as well by optimising the coupling between the bus waveguide and MRR cavity~\cite{Vernon16}. These results, recent demonstrations of quantum frequency conversion~\cite{Li2016}, and CMOS compatible fabrication represent a promising path forward for $ \rm Si_3 N_4$ as a key enabling quantum technologies. 

%

\section{Acknowledgments}
The authors would like to thank M. Perrenoud, M. Caloz and F. Bussi\`{e}res for development of the SNSPDs. This work was supported by the Swiss National Science Foundation (Grant No. 200020\_182664 and No. 161573 (precoR)). This publication was supported by Contract HR0011-15-C-0055 (DODOS) from the Defense Advanced Research Projects Agency (DARPA) and the  Microsystems Technology Office (MTO). M. Karpov acknowledges the support from the European Space Technology Centre with ESA Contract No. 4000116145/16/NL/MH/GM.
 
\bibliography{references}

\end{document}